\documentclass[amsfonts,aps,nofootinbib,preprint,superscriptaddress]{revtex4}

\usepackage{graphicx}

\begin{document}

\title{Superspace approach to the renormalization of the O'Raifeartaigh model up to the second order in the LDE parameter}

\author{M. C. B. Abdalla{\footnote{mabdalla@ift.unesp.br}}}
\affiliation{Instituto de F\'{\i}sica Te\'{o}rica, UNESP - Universidade Estadual Paulista, Rua Dr. Bento Teobaldo Ferraz 271, 
Bloco II, Barra-Funda, Caixa Postal 70532-2, 01156-970, S\~{a}o Paulo, SP, Brazil }
\author{J. A. Helay\"{e}l-Neto{\footnote{helayel@cbpf.br}}}
\affiliation{Centro Brasileiro de Pesquisas F\'isicas, Rua Dr. Xavier Sigaud 150, 
Urca, Rio de Janeiro, RJ, 22290-180, Brazil}
\author{Daniel L. Nedel{\footnote{daniel.nedel@unipampa.edu.br}}}
\affiliation{Universidade Federal do Pampa, 
Rua Carlos Barbosa S/N, Bairro Get\'ulio Vargas, 96412-420, Bag\'e, RS, Brazil}
\author{Carlos R. Senise Jr.{\footnote{carlossenise@unipampa.edu.br}}}
\affiliation{Universidade Federal do Pampa, Av. Pedro Anuncia\c{c}\~ao S/N, Vila Batista, 96570-000, Ca\c{c}apava do Sul, RS, Brazil}

\begin{abstract}
We adopt a superspace/supergraph formalism to pursue the investigation of the structure of one- and two-loop divergences in the frame of the minimal O'Raifeartaigh model that realizes the F-term spontaneous supersymmetry breaking. The linear delta expansion(LDE) procedure is introduced and renormalization is carried out up to the second order in the LDE expansion parameter. In agreement with the nonrenormalization theorem for the (chiral/antichiral) matter potential of ${\mathcal N}=1, \ D=4$ supersymmetry, our explicit supergraph calculations confirm that only the K\"{a}hler potential is actually renomalized.
\end{abstract}

\maketitle

Supersymmetry (SUSY) and  viable procedures to investigate its explicit, spontaneous and dynamical breaking mechanisms \cite{Zum} are topics of constant and renewed interest in the literature, in view of the structural r\^ole SUSY plays in the construction of field-theoretic frameworks for fundamental interactions and realistic models for Elementary Particle Physics. In the early days of the dawn of SUSY in high energy particle theories, it was realized that spontaneous breakdown of this new fermion/boson symmetry was a topic of major relevance in order to make contact between SUSY and
observations and communicate high-energy SUSY imprints with the low energy sector of the Standard Model. Ever since, in different scenarios like the MSSM \cite {Dim}, Kaluza-Klein Supergravities \cite{Duff}, string inspired models for elementary particle interactions \cite{Lopes}, Seiberg-Witten super-Yang-Mills dualities \cite{SW} and brane-world scenarios \cite{Luty}, the problem of SUSY breaking has been suitably reassessed. More recently, SUSY breaking in connection with brane physics in general and M2-brane modeling \cite{BL}, more specifically,  has triggered a great deal of attention to the understanding of a number of issues related to (2+1)-dimensional supersymmetric field theories \cite{Buch1}, like the so-called ABJM models \cite{ABJ}.

If, on the one hand, SUSY breaking mechanisms and their consequences are relevant for connecting high energy fundamental physics to the regime of accelerator energies and for establishing the consistency of more formal field theoretic models, on the other hand, one must develop technical methods to suitably carry out the SUSY breaking program  and to pursue its investigation perturbatively and, hopefully, by means of some nonperturbative scheme. In this case, if it is not possible to derive an exact result, one could somehow come over this problem by devising some sort of semi-perturbative scheme based upon the resummation of a certain class of (perturbative) Feynman diagrams and attain, thereby, a result that incorporates all orders
in some coupling parameter.

In the frame  of perturbative and ressumation methods, the calculation and study of the effective potential is a viable path to be followed. In this context, different methods for effective potential calculation have been proposed in the literature soon after the idea of spontaneous symmetry breaking and the Higgs mechanism were adopted to realize the breakdown of gauge symmetries and to introduce the hierarchy of energy scales in unified models for elementary particle interactions. Here, we shall be concentrating our efforts to apply the so called linear delta expansion (LDE) \cite{OPT}, suitably extended \cite{WZdelta,OFTdelta} to incorporate superspace and superfield techniques, to compute corrections at the one- and two-loop orders to the effective potential of the minimal O'Raifeartaigh model \cite{O'Raifeartaigh} which spontaneously breaks ${\mathcal N}=1$ SUSY in four space-time dimensions. We stress that, even though SUSY is broken, a superfield approach is still the most appropriate setup to describe the problem.  The main characteristic of the LDE  is to use a traditional perturbative approach together with an optimization procedure. So, in order to derive a result in all orders of the coupling constant, it is just necessary to work with a few diagrams and use perturbative renormalization techniques.

Having in mind the ever increasing importance of understanding and proposing new scenarios for SUSY breaking, our paper sets out to tackle a specific problem - the application of the LDE at the second order to the minimal O'Raifeartaigh model - mainly motivated by the reasons that follow below:

(i) to exploit superfield techniques and superspace methods in connection with the LDE procedure to compute higher-order corrections to the effective potential of a supersymmetric field model, even if SUSY is spontaneously broken and we are obliged to deal with terms that explicitly break SUSY in  superspace. This may show us nontrivial technicalities and features whenever we insist in performing superfield calculations to describe SUSY breaking. Our viewpoint is that superspace, with its corresponding tensor calculus expressed in terms of superfields, is still the most suitable tool to deal with even if SUSY is lost;

(ii) to use the supergraph approach to carry out  the superspace renormalization of supersymmetric models if  one adopts the LDE procedure to compute the loop corrected effective potential. In the present paper, we shall be concerned with the structure of divergences and the renormalization of the model we pick out to work with. The optimization procedure and the attainment of the full (LDE) two-loop corrected effective potential shall be reported on in a forthcoming work \cite{nos2};

(iii) once the whole treatment is understood for this more traditional case of F-type breaking, we shall be able to apply it to other interesting situations, such as SUSY breaking by a D-term in supersymmetric gauge theories, \textit{R} symmetry spontaneous breakdown in connection with the existence of metastable SUSY breaking vacua \cite {metastable} and the computation of loop quantum contributions to the effective K\"{a}hler and chiral potentials \cite{Buch2}.

Keeping in mind this whole framework and the motivations mentioned above, we organize the present paper according to the following outline: in Section II, we recall the main features of the LDE in superspace, highlighting the superfield techniques in the presence of terms that explicitly break SUSY. Section III is devoted to report and discuss the superspace evaluation of the results attained at order one in the $\delta$ parameter; next, in Section IV, we go  a step further and analyze the divergent structure of the order two contributions, which encompass one- and two-loop supergraphs, to show how to renormalize the model to the first and second order in the parameter of LDE. With our calculations, we shall see that only the K\"{a}hler potential is actually renormalized, as expected by the nonrenormalization of the chiral potential in ${\mathcal N}=1, \ D=4$ SUSY. Finally, our Concluding Remarks are cast in Section V. Two Appendices follow: in Appendix A, we work out the main results in connection with the integration over the Grassmannian sector of superspace; in Appendix B, we present the explicit answers to the momentum-space loop integrals that appear in the course of our calculations.

          
\section{The Linear Delta Expansion in Superspace}
In this section, we are going to make a brief review of the application of the linear delta expansion to supersymmetric theories. We follow the references \cite{WZdelta,OFTdelta}. Starting with a Lagrangian ${\cal L}$, let us define the following interpolated Lagrangian ${\cal L}^{\delta}$:
\begin{equation}
{\mathcal L}^{\delta}=\delta{\mathcal L}(\mu)+(1-\delta){\mathcal L}_{0}(\mu) \ , \label{LDE} 
\end{equation}
where $\delta$ is an arbitrary parameter, ${\cal L}_0(\mu)$ is the free Lagrangian, and $\mu$ is a mass parameter. Note that, when $\delta =1$, the original theory is retrieved. The $\delta$ parameter labels interactions and is used as a perturbative coupling instead of the original one. The mass parameter appears in ${\cal L}_0$ and $\delta{\cal L}_0$. The $\mu$ dependence of ${\cal L}_0$ is absorbed into the propagators, whereas $\delta{\cal L}_0$ is regarded as a quadratic interaction. 
 
Let us now define the strategy of the method. We apply an usual perturbative expansion in $\delta$ and, at the end of the calculation, we set $\delta=1$. Up to this stage, traditional perturbation theory is applied, working with finite Feynman diagrams, and the results are purely perturbative. However, quantities evaluated at finite order in $\delta$ explicitly depend on $\mu$. So it is necessary to fix the $\mu$ parameter. There are two ways to do that. The first one is to use the principle of minimal sensitivity (PMS) \cite{PMS}. It requires  that a physical quantity, such as the effective potential $V^{(k)}(\mu)$, calculated perturbatively to order $\delta ^k$, must be evaluated at a point where it is less sensitive to the parameter $\mu$. According to the PMS, $\mu={\mu_0}$ is the solution to the equation
\begin{equation}
\left.\frac{\partial V^{(k)}(\mu)}{\partial\mu}\right|_{\mu=\mu_0,\delta=1}=0 \ . \label{PMS}
\end{equation}
After this procedure, the optimum value, ${\mu_0}$, will be a function of the original coupling and fields. Then, we replace ${\mu_0}$ into the effective potential $V^{(k)}$ and obtain a nonperturbative result, since the propagator depends on $\mu$. 
 
The second way to fix $\mu$ is known as the fastest apparent convergence (FAC) criterion \cite{PMS}. It requires that, for any $k$ coefficient of the perturbative expansion 
\begin{equation}
V^{(k)}(\mu)=\sum_{i=0}^{k}c_i(\mu)\delta^i \label{coef} \ ,
\end{equation}
the following relation must be fulfilled:   
\begin{equation}
\left.\left[V^{(k)}(\mu)- V^{(k-1)}(\mu)\right]\right|_{\delta=1}=0 \ . \label{fac}
\end{equation}
 
Again, the ${\mu_0}$ solution of the above equation will be a function of the original couplings and fields, and whenever we replace $\mu={\mu_0}$ into $V(\mu)$, we obtain a nonperturbative result. Equation (\ref{fac}) is equivalent to taking the $k$th coefficient of (\ref{coef}) equal to zero ($c_k=0$). If we are interested in an order-$\delta^{k}$ result [$V^{(k)}(\mu)$] using the FAC criterion, it is just necessary to find the solution to the equation $\displaystyle\left.c_{k+1}(\mu)\right|_{\mu=\mu_0}=0$ and plug it into $V^{(k)}(\mu)$. References \cite{Kneur,Ricardo}
provide an extensive list of successful applications of the method. 

Let us now further develop the LDE for superspace applications. Following Ref. \cite{WZdelta}, for general models with chiral and antichiral superfields, we need to implement two mass parameters, $\mu$ and $\bar{\mu}$, instead of just one. In order to fix these parameters, we employ two optimization equations. Also, we need to take care of the vacuum diagrams. In general, when the effective potential is calculated in quantum field theory, we do not worry about vacuum diagrams, since they do not depend on fields. However, the vacuum diagrams depend on $\mu$ and are important to the LDE, since the arbitrary mass parameter will depend on fields after the optimization procedure. So, in the LDE, it is necessary to calculate the vacuum diagrams order by order. On the other hand, it is well-known that, in superspace, vacuum superdiagrams are identically zero, by virtue of Berezin integrals. To avoid this, we have to consider, from the very beginning, the parameters $\mu$, $\bar{\mu}$ as superfields and keep the vacuum supergraphs until the optimization procedure is carried out. In order to make the procedure clear, let us write the interpolated Lagrangian, ${\cal L}^\delta$, for the Wess-Zumino model discussed in \cite{WZdelta}:
\begin{eqnarray}
{\mathcal L}^{\delta}&=&\delta{\mathcal L}(\mu,\bar{\mu})+(1-\delta){\mathcal L}_{0}(\mu,\bar{\mu}) \nonumber\\
&=&\int\!d^{4}\theta\bar{\Phi}\Phi+\!\int\!d^{2}\theta\left(\frac{M}{2}\Phi^{2}+\frac{\delta\lambda}{3!}\Phi^{3}-\frac{\delta\mu}{2}\Phi^{2}\right)+\!\int\!d^{2}\bar{\theta}\left(\frac{\bar{M}}{2}\bar{\Phi}^{2}+\frac{\delta\bar{\lambda}}{3!}\bar{\Phi}^{3}-\frac{\delta\bar{\mu}}{2}\bar{\Phi}^{2}\right),  \label{linter}
\end{eqnarray} 
where $m$ is the original mass, $M=m+\mu$ and $\bar{M}=m+\bar{\mu}$. Now, one has a new chiral and antichiral quadratic interaction proportional to $\delta\mu$ and $\delta\bar{\mu}$. Also the superpropagator will have a dependence on $\mu$ and $\bar{\mu}$.
From the generating superfunctional in the presence of the chiral ($J$) and antichiral ($\bar J$) sources
\begin{equation}
\tilde{Z}[J,\bar{J}]=exp\left[iS_{INT}\left(\frac{1}{i}\frac{\delta}{\delta J},\frac{1}{i}\frac{\delta}{\delta\bar{J}}\right)\right]exp\left[\frac{i}{2}
(J,\bar{J})G^{(M,\bar{M})}\left( \begin{array}{c}
J \\
\bar{J}
\end{array}
\right)\right] \ ,
\end{equation}
we can write the supereffective action:
\begin{equation}
\Gamma[\Phi,\bar{\Phi}]=-\frac{i}{2}\ln[sDet\left(G^{(M,\bar{M})}\right)]-i\ln\tilde{Z}[J,\bar{J}]-\int\!d^{6}zJ(z)\Phi(z)-\int\!d^{6}\bar{z}\bar{J}(z)\bar{\Phi}(z),             \label{sea}
\end{equation}
where $G^{(M,\bar{M})}$ is the matrix propagator and $sDet\left(G^{(M,\bar{M})}\right)$ is the superdeterminant of $G^{(M,\bar{M})}$, which, in general, is equal to one; but here we keep it, because $G^{(M,\bar{M})}$ depends on $\mu$ and $\bar{\mu}$. Also, due to the $\mu$ and $\bar{\mu}$ dependence, the supergenerator of the vacuum diagrams, $\tilde{Z}[0,0]$, is not identically equal to one. We can define the normalized functional generator as $Z_N = \frac{\tilde{Z}[J,\bar J]}{\tilde{Z}[0,0]}$, and write the effective action as
\begin{equation}
\Gamma[\Phi,\bar{\Phi}]=-\frac{i}{2}\ln[sDet(G)]-i\ln\tilde{Z}[J_{0},\bar{J}_{0}]+\Gamma_{N}[\Phi,\bar{\Phi}] \ , \label{Gamma Phi barPhi}
\end{equation}
where the sources $J_0$ and $\bar{J}_0$ are defined by the equations 
\begin{eqnarray}
\frac{\delta W[J,\bar{J}]}{\delta J(z)}|_{J=J_{0}}=\frac{\delta W[J,\bar{J}]}{\delta \bar{J}(z)}|_{\bar{J}=\bar{J}_{0}}=\frac{\delta\tilde{Z}[J,\bar{J}]}{\delta J(z)}|_{J=J_{0}}=\frac{\delta\tilde{Z}[J,\bar{J}]}{\delta\bar{J}(z)}|_{\bar{J}=\bar{J}_{0}}=0 \ . \label{gerador}
\end{eqnarray}
In (\ref{Gamma Phi barPhi}), the first two terms represent the vacuum diagrams (which are usually zero) and $\Gamma_N[\Phi,\bar{\Phi}]$ is the usual contribution to the effective action. 
 
Let us now derive the interpolated Lagrangian and the new Feynman rules for the O'Raifeartaigh model. The simplest O'Raifeartaigh model is described by the following Lagrangian:
\begin{equation}
{\mathcal L}=\int d^{4}\theta\bar{\Phi}_{i}\Phi_{i}-\left[\int d^{2}\theta\left(\xi\Phi_{0}+m\Phi_{1}\Phi_{2}+g\Phi_{0}\Phi_{1}^{2}\right)+h.c.\right] \ , \label{O'Raifeartaigh}
\end{equation} 
where $i=0,1,2$.
  
Following reference \cite{OFTdelta}, in order to take into account the nonperturbative contributions of all fields of the model, we need to implement the LDE with the matrix mass parameters $\mu_{ij}$ and $\bar{\mu}_{ij}$. Adding and subtracting these mass terms in the Lagrangian of a general O'Raifeartaigh model we obtain
\begin{equation}
{\mathcal L}(\mu,\bar{\mu})={\mathcal L}_{0}(\mu,\bar{\mu})+{\mathcal L}_{int}(\mu,\bar{\mu}) \ ,
\end{equation}
where
\begin{eqnarray} 
{\mathcal L}_{0}(\mu,\bar{\mu})&=&\int d^{4}\theta\bar{\Phi}_{i}\Phi_{i}-\left[\int d^{2}\theta\left(\xi_{i}\Phi_{i}+\frac{1}{2}M_{ij}\Phi_{i}\Phi_{j}\right)+h.c.\right] \ , \\
{\mathcal L}_{int}(\mu,\bar{\mu})&=&-\left[\int d^{2}\theta\left(\frac{1}{3!}g_{ijk}\Phi_{i}\Phi_{j}\Phi_{k}-\frac{1}{2}\mu_{ij}\Phi_{i}\Phi_{j}\right)+h.c.\right] \ ,  
\end{eqnarray}
with $M_{ij}=m_{ij}+ \mu_{ij}$ and $i,j,k=0,1,2$ are symmetrical indices.



Let us expand the arbitrary mass parameters as chiral and antichiral superfields:
\begin{equation}
\mu_{ij}=\lambda_{ijk}\varphi_k=\lambda_{ijk}(\rho_k+\theta^2\chi_k)=\lambda_{ijk}\rho_k+\lambda_{ijk}\chi_k\theta^2=\rho_{ij}+b_{ij}\theta^2 \ ,
\end{equation}
so that
\begin{equation}
M_{ij}=m_{ij}+\mu_{ij}=(m_{ij}+\rho_{ij})+b_{ij}\theta^2=a_{ij}+b_{ij}\theta^2 \ .
\end{equation}

Now, the interpolated Lagrangian (\ref{LDE}) becomes
\begin{equation}
{\mathcal L}^{\delta}= {\mathcal L}_{0}^{\delta} + {\mathcal L}_{int}^{\delta} \ ,
\end{equation}
where the free Lagrangian, ${\mathcal L}^{\delta}_0$, is 
\begin{equation}
{\mathcal L}_{0}^{\delta}=\int d^{4}\theta\bar{\Phi}_{i}\Phi_{i}-\left[\int d^{2}\theta\left(\xi_{i}\Phi_{i}+\frac{1}{2}a_{ij}\Phi_{i}\Phi_{j}+\frac{1}{2}b_{ij}\theta^{2}\Phi_{i}\Phi_{j}\right)+h.c.\right] \ ,
\end{equation}
and the interaction Lagrangian reads as follows:
\begin{equation}
{\mathcal L}_{int}^{\delta}=-\left[\int d^{2}\theta\left(\frac{\delta}{3!}g_{ijk}\Phi_{i}\Phi_{j}\Phi_{k}-\frac{\delta}{2}\mu_{ij}\Phi_{i}\Phi_{j}\right)+h.c.\right] \ .
\end{equation}

Notice that the interaction Lagrangian has now soft breaking terms proportional to the $\mu$ components.  We are going to treat these terms perturbatively in $\delta$, like all interactions. 

Now, in order to get the simplest O'Raifeartaigh model when $\delta=1$ (\ref{O'Raifeartaigh}), we make the choices
\begin{equation}
\left\{\begin{array}{lllll}
          \xi_{0}=\xi \ ; \\
          M_{01}=a_{01}=\rho_{01}=a \ ; \\
          M_{11}=b_{11}\theta^{2}=b\theta^{2} \ ; \\
          M_{12}=a_{12}=m_{12}+\rho_{12}=m+\rho=M \ ; \\
          g_{011}=g \ ,
          \end{array}\right. \label{choice}
\end{equation}
and all other $\xi_{i}$ and $M_{ij}$ set to zero. With that, we obtain
\begin{eqnarray}
{\mathcal L}_{0}^{\delta}&=&\int d^{4}\theta\bar{\Phi}_{i}\Phi_{i}-\left[\int d^{2}\theta\left(\xi\Phi_{0}+M\Phi_{1}\Phi_{2}+a\Phi_{0}\Phi_{1}+\frac{1}{2}b\theta^{2}\Phi_{1}^{2}\right)+h.c.\right] \ , \nonumber\\
{\mathcal L}_{int}^{\delta}&=&-\left[\int d^{2}\theta\left(\delta g\Phi_{0}\Phi_{1}^{2}-\delta\rho\Phi_{1}\Phi_{2}-\delta a\Phi_{0}\Phi_{1}-\frac{\delta}{2}b\theta^{2}\Phi_{1}^{2}\right)+h.c.\right] \ .
\end{eqnarray}

As is well-known, this O'Raifeartaigh model has an {\it R} symmetry. The {\it R} charges of quiral superfields $\Phi_0$, $\Phi_1$, $\Phi_2$ are  respectively $R_{0}=2$, $R_{1}=0$ and $R_{2}=2$. In order to preserve the {\it R} symmetry in the interpolated Lagrangian, the {\it R} charges of the parameters $a$ and $b$ are $R_{a}=0$ and $R_{b}=0$, which must be preserved after the optimization procedure. 

The new propagators can be derived from the free Lagrangian, which also has an explicit dependence on $\theta$ and $\bar{\theta}$ from the $\mu$ and $\bar{\mu}$ components. Using the techniques developed in \cite{Helayel}, the propagators can be written as 
\begin{eqnarray}
\langle\Phi_{0}\bar{\Phi}_{0}\rangle&=&(k^{2}+|M|^{2})A(k)\delta^{4}_{12}+|a|^{2}|b|^{2}B(k)\theta_{1}^{2}\bar{\theta}_{1}^{2}\delta_{12}^{4} \ ; \nonumber\\
\langle\Phi_{0}\bar{\Phi}_{1}\rangle&=&\bar{a}bC(k)\frac{1}{16}D_{1}^{2}\bar{D}_{1}^{2}\theta_{1}^{2}\delta_{12}^{4} \ ; \nonumber\\
\langle\Phi_{0}\bar{\Phi}_{2}\rangle&=&-M\bar{a}A(k)\delta^{4}_{12}+M\bar{a}|b|^{2}B(k)\theta_{1}^{2}\bar{\theta}_{1}^{2}\delta^{4}_{12} \ ; \nonumber\\
\langle\Phi_{1}\bar{\Phi}_{1}\rangle&=&E(k)\delta_{12}^{4}+|b|^{2}B(k)\frac{1}{16}D_{1}^{2}\theta_{1}^{2}\bar{\theta}_{1}^{2}\bar{D}_{1}^{2}\delta_{12}^{4} \ ; \nonumber\\
\langle\Phi_{1}\bar{\Phi}_{2}\rangle&=&-M\bar{b}F(k)\bar{\theta}_{1}^{2}\delta^{4}_{12} \ ; \nonumber\\
\langle\Phi_{2}\bar{\Phi}_{2}\rangle&=&(k^{2}+|a|^{2})A(k)\delta^{4}_{12}+|M|^{2}|b|^{2}B(k)\theta_{1}^{2}\bar{\theta}_{1}^{2}\delta_{12}^{4} \ ; \nonumber\\
\langle\Phi_{0}\Phi_{0}\rangle&=&-|a|^{2}\bar{b}C(k)\frac{1}{4}D_{1}^{2}\theta_{1}^{2}\delta_{12}^{4} \ ; \nonumber\\
\langle\Phi_{0}\Phi_{1}\rangle&=&\bar{a}A(k)\frac{1}{4}D_{1}^{2}\delta_{12}^{4}-\bar{a}|b|^{2}B(k)\frac{1}{4}\theta_{1}^{2}\bar{\theta}_{1}^{2}D_{1}^{2}\delta_{12}^{4} \ ; \nonumber\\
\langle\Phi_{0}\Phi_{2}\rangle&=&-\bar{M}\bar{a}\bar{b}C(k)\frac{1}{4}D_{1}^{2}\theta_{1}^{2}\delta^{4}_{12} \ ; \nonumber\\
\langle\Phi_{1}\Phi_{1}\rangle&=&\bar{b}F(k)\frac{1}{4}\bar{\theta}_{1}^{2}D_{1}^{2}\delta_{12}^{4} \ ; \nonumber\\
\langle\Phi_{1}\Phi_{2}\rangle&=&\bar{M}A(k)\frac{1}{4}D_{1}^{2}\delta^{4}_{12}-\bar{M}|b|^{2}B(k)\frac{1}{4}D_{1}^{2}\theta_{1}^{2}\bar{\theta}_{1}^{2}\delta^{4}_{12} \ ; \nonumber\\
\langle\Phi_{2}\Phi_{2}\rangle&=&-|M|^{2}\bar{b}C(k)\frac{1}{4}D_{1}^{2}\theta_{1}^{2}\delta^{4}_{12} \ , \label{propagators}
\end{eqnarray}
with
\begin{eqnarray*}
A(k)&=&\frac{1}{k^{2}\left(k^{2}+|M|^{2}+|a|^{2}\right)} \ \ , \\ B(k)&=&\frac{1}{\left(k^{2}+|M|^{2}+|a|^{2}\right)\left[\left(k^{2}+|M|^{2}+|a|^{2}\right)^{2}-|b|^{2}\right]} \ \ , \\
C(k)&=&\frac{1}{k^{2}\left[\left(k^{2}+|M|^{2}+|a|^{2}\right)^{2}-|b|^{2}\right]} \ \ , \\ 
E(k)&=&\frac{1}{k^{2}+|M|^{2}+|a|^{2}} \ \ , \\ 
F(k)&=&\frac{1}{\left(k^{2}+|M|^{2}+|a|^{2}\right)^{2}-|b|^{2}} \ \ .
\end{eqnarray*} 

In usual quantum field theories, the optimized parameters appear at the poles of the propagators, as mass terms.
Here, it should be emphasized the nontrivial dependence on the parameters $a$, $b$ and $\rho$, which appear not only at the poles, but also in the numerators.     

We can also write the new Feynman rules for the vertices:
\begin{eqnarray}
\Phi_{0}\Phi_{1}^{2} \ \ \hbox{vertex}&:&2\delta g\int d^{4}\theta \ ; \nonumber\\
\Phi_{1}\Phi_{2} \ \ \hbox{vertex}&:&-\delta\rho\int d^{4}\theta \ ; \nonumber\\
\Phi_{0}\Phi_{1} \ \ \hbox{vertex}&:&-\delta a\int d^{4}\theta \ ; \nonumber\\
\Phi_{1}\Phi_{1} \ \ \hbox{vertex}&:&-\frac{\delta b}{2}\int d^{4}\theta\theta^{2} \ . \label{vertices}
\end{eqnarray}

Now we have all the necessary ingredients to calculate the effective potencial using pertubation theory in $\delta$. In the next section we show the order one results.  

\section{Order one results}

The perturbative effective potential can now be calculated in powers of $\delta$ using the one particle irreducible functions, defined in the expansion of the effective action, taking into account vacuum diagrams. In reference \cite{OFTdelta}, it was shown that, after the optimization procedure at order one, the optmized effective potential provides the sum of all one-loop diagrams. In that case, analytical solutions were obtained for the optimization procedure before calculate the superspace and momentum integrals. However, in order to go beyond the one-loop approximation, it is necessary to go beyond the order $\delta^1$. In this case, it is not possible to find solutions before evaluating the superspace and momentum integrals and the optimization procedure must be carried out after the renormalization of the theory. Owing to the nontrivial dependence of the propagators on the optimized parameters, it is not clear that the method does not alter the divergences structure of the model. We investigate this fact here.  

In Fig. 1, one can see the diagrammatic sum of the effective potential up to the order $\delta^{1}$ (${\mathcal V}_{eff}^{(1)}$).
\begin{eqnarray*}
\begin{picture}(350,5) \thicklines
\put(15,0){\circle{25}}\put(30,-3){+}
\put(55,0){\circle{25}}\put(67,0){\line(50,0){20}}\put(73,4){$\Phi_0$}\put(91,-3){+}
\put(116,0){\circle{25}}\put(123,-3){$\times$}\put(132,0){$\theta^2$}\put(120,14){$\Phi_1$}\put(120,-22){$\Phi_1$}\put(145,-3){+}
\put(171,0){\circle{25}}\put(183,0){\line(50,0){20}}\put(189,4){$\Phi_1$}\put(208,-3){+}
\put(233,0){\circle{25}}\put(240.5,-3){$\times$}\put(237,14){$\Phi_0$}\put(237,-22){$\Phi_1$}\put(255,-3){+}
\put(282,0){\circle{25}}\put(289,-3){$\times$}\put(286,14){$\Phi_1$}\put(286,-22){$\Phi_2$}\put(304,-3){+}
\put(321,-3){$h.c.$}
\end{picture}
\end{eqnarray*}

\vspace{0,1cm}
\begin{center}
Fig. 1: Effective potential up to the order $\delta^1$. 
\end{center}

Note that, by virtue of the $\theta$-dependent propagators, the tadpole diagrams are not identically zero, as  usual in superspace. The first diagram is of order $\delta^{0}$ and corresponds to the first term of the effective action expansion, defined in (\ref{Gamma Phi barPhi}). 

Using the Feynman rules and the results of \cite{OFTdelta}, the expressions for the diagrams of the figure above, in the order they appear are:
\begin{itemize}
\item Order $\delta^0$ (vacuum diagram):
\begin{equation} 
{\mathcal G}^{(1)\delta^{0}}=\frac{1}{2}\int d^{4}\theta_{12}\delta^{4}_{21}Tr\ln[P^{T}K]\delta^{4}_{21} \ , \label{ordem 0}
\end{equation}
where $d^{4}\theta_{12}=d^{4}\theta_{1}d^{4}\theta_{2}$ and  $Tr$ is  the trace over the quiral multiplets defined in the real basis by $(\Phi^T,\bar{\Phi})^T$. Details of this calculation can be seen in ref. \cite{OFTdelta, Nibbelink}. The matrix $P$ is defined by the chiral projectors $P_{+}=\frac{\bar{D}^{2}D^{2}}{16\Box}$ and $P_{-}=\frac{D^{2}\bar{D}^{2}}{16\Box}$ as
\begin{equation}
P=\left(\begin{array}{cc}
          0 & P_{-} \\
          P_{+} & 0
          \end{array}\right) \ , \label{P}
\end{equation}
and
\begin{equation}
K=\left(\begin{array}{cc}
          \left(AP_{-}+B\frac{1}{\Box^{1/2}}\eta_{-}\right)\frac{D^2}{4\Box} & \textbf{1}_{3} \\
          \textbf{1}_{3} & \left(\bar{A}P_{+}+\bar{B}\frac{1}{\Box^{1/2}}\bar{\eta}_{+}\right)\frac{\bar{D}^2}{4\Box}
          \end{array}\right) \ , \label{K}
\end{equation}
with
\begin{equation}
A=\left(\begin{array}{ccc}
          0 & a & 0 \\
          a & 0 & M \\
          0 & M & 0
          \end{array}\right) \ , \ 
B=\left(\begin{array}{ccc}
          0 & 0 & 0 \\
          0 & b & 0 \\
          0 & 0 & 0
          \end{array}\right) \ , \ 
\eta_{-}=\Box^{1/2}P_{-}\theta^{2}P_{-} \ , \ 
\bar{\eta}_{+}=\Box^{1/2}P_{+}\bar{\theta}^{2}P_{+} \ , \label{matrizes}                   
\end{equation}
is the quadratic part of the free Lagrangian (there are also  quadratic terms in the  interaction Lagrangian, wich depend on the optimization parameters). Using the results of \cite{OFTdelta}, the eq. (\ref{ordem 0}) can be written as:
\begin{equation}
{\mathcal G}^{(1)\delta^0}=\frac{1}{2}\int\frac{d^{4}k}{(2\pi)^{4}}\ln\left[1-\frac{\left|b\right|^{2}}{\left(k^{2}+\left|M\right|^{2}+\left|a\right|^{2}\right)^{2}}\right] \ . \label{ln}
\end{equation}\item Order $\delta^1$:
\begin{eqnarray}
{\mathcal G}_{1}^{(1)\delta^{1}}&=&-2\delta g\bar{b}\int\frac{d^{4}k}{(2\pi)^{4}}F(k)\int d^{4}\theta\bar{\theta}^{2}\Phi_{0}+h.c. \ ; \nonumber\\
{\mathcal G}_{2}^{(1)\delta^{1}}&=&\frac{1}{2}\delta|b|^{2}\int\frac{d^{4}k}{(2\pi)^{4}}F(k)\int d^{4}\theta\theta^{2}\bar{\theta}^{2}+h.c. \ ; \nonumber\\
{\mathcal G}_{3}^{(1)\delta^{1}}&=&4\delta g\bar{a}|b|^{2}\int\frac{d^{4}k}{(2\pi)^{4}}B(k)\int d^{4}\theta\theta^{2}\bar{\theta}^{2}\Phi_{1}+h.c. \ ; \nonumber\\
{\mathcal G}_{4}^{(1)\delta^{1}}&=&-\delta|a|^{2}|b|^{2}\int\frac{d^{4}k}{(2\pi)^{4}}B(k)\int d^{4}\theta\theta^{2}\bar{\theta}^{2}+h.c. \ ; \nonumber\\
{\mathcal G}_{5}^{(1)\delta^{1}}&=&-\delta\rho\bar{M}|b|^{2}\int\frac{d^{4}k}{(2\pi)^{4}}B(k)\int d^{4}\theta\theta^{2}\bar{\theta}^{2}+h.c. \ . \label{ordem 1}
\end{eqnarray}
\end{itemize}

Note that all the expressions for the superdiagrams were written as an integral over $d^{4}\theta$, according to the nonrenormalization theorem. In Appendix A we give the final results for the superspace integrals. Using the results
\begin{equation}
\int d^{4}\theta\bar{\theta}^{2}\Phi_{0}=\int d^{2}\theta\Phi_{0} \ \ \ , \ \ \ \int d^{4}\theta\theta^{2}\bar{\theta}^{2}=1 \ \ \ \mbox{e} \ \ \ \int d^{4}\theta\theta^{2}\bar{\theta}^{2}\Phi_{1}=\int d^{2}\theta\theta^{2}\Phi_{1} \ , \label{rel int superspace} 
\end{equation}
the effective potential up to the order $\delta^1$ is given by:
\begin{eqnarray}
{\mathcal V}_{eff}^{(1)}&=&{\mathcal G}^{(1)\delta^{0}}+\sum_{i=1}^{5}{\mathcal G}_{i}^{(1)\delta^{1}} \nonumber\\
&=&\frac{1}{2}\int d^{4}\theta_{12}\delta^{4}_{12}Tr\ln[P^{T}K]\delta^{4}_{12}+\delta\int\frac{d^{4}k}{(2\pi)^{4}}F(k)\left\{-2g\bar{b}\int d^{2}\theta\Phi_{0}+\frac{1}{2}\left|b\right|^{2}+h.c.\right\} \nonumber\\
&&+\delta\int\frac{d^{4}k}{(2\pi)^{4}}B(k)\left\{4g\bar{a}\left|b\right|^{2}\int d^{2}\theta\theta^{2}\Phi_{1}-\left|a\right|^{2}\left|b\right|^{2}-\rho\bar{M}\left|b\right|^{2}+h.c.\right\} \ . \label{pot efet OR ordem 1}
\end{eqnarray} 

Now we are going to regularize  the integrals that appear in the effective potential up to the order $\delta^1$.  
The diagrams of order $\delta^1$ are shown in Fig 2.
\begin{eqnarray*}
\begin{picture}(375,5) \thicklines  
\put(55,0){\circle{25}}\put(67,0){\line(50,0){20}}\put(73,4){$\Phi_0$}\put(91,-3){+}
\put(116,0){\circle{25}}\put(123,-3){$\times$}\put(132,0){$\theta^2$}\put(120,14){$\Phi_1$}\put(120,-22){$\Phi_1$}\put(145,-3){+}
\put(171,0){\circle{25}}\put(183,0){\line(50,0){20}}\put(189,4){$\Phi_1$}\put(208,-3){+}
\put(233,0){\circle{25}}\put(240.5,-3){$\times$}\put(237,14){$\Phi_0$}\put(237,-22){$\Phi_1$}\put(255,-3){+}
\put(282,0){\circle{25}}\put(289,-3){$\times$}\put(286,14){$\Phi_1$}\put(286,-22){$\Phi_2$}\put(304,-3){+}
\put(321,-3){$h.c. \ ,$}
\end{picture}
\end{eqnarray*}
\vspace{-0,4cm}
\begin{center}
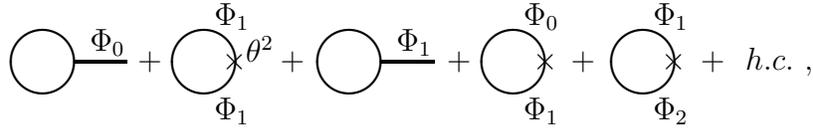
Fig. 2: One-loop diagrams of order $\delta^1$. 
\end{center}

We are going to use the notation defined in (\ref{etas}). 
Using the results of  Appendix B, these one-loop diagrams are written as:
\begin{eqnarray}
{\mathcal G}_{1}^{(1)\delta^{1}}&=&2\delta g\int\frac{d^{4}kd^{4}\theta_{12}}{(2\pi)^4}\Phi_{0}(1)\delta^{4}_{12}\left[-\frac{1}{4}\bar{D}_{1}^{2}(k)\langle\Phi_{1}\Phi_{1}\rangle\right]+h.c. \nonumber\\
&=&-2\delta g\bar{b}\int\frac{d^{4}k}{(2\pi)^{4}}F(k)\int d^{4}\theta\bar{\theta}^{2}\Phi_{0}+h.c. \nonumber\\
&=&-\frac{4\delta gb\langle F_{0}\rangle}{\kappa}\frac{1}{\epsilon}+\frac{2\delta g\langle F_{0}\rangle}{\kappa}\left(\eta^{+}\overline{\ln}\eta^{+}-\eta^{-}\overline{\ln}\eta^{-}-2b\right) \ ; \label{G1 delta1}
\end{eqnarray}
\begin{eqnarray}
{\mathcal G}_{2}^{(1)\delta^{1}}&=&-\frac{\delta b}{2}\int\frac{d^{4}kd^{4}\theta_{12}}{(2\pi)^4}\theta_{1}^{2}\delta^{4}_{12}\left[-\frac{1}{4}\bar{D}_{1}^{2}(k)\langle\Phi_{1}\Phi_{1}\rangle\right]+h.c. \nonumber\\
&=&\frac{\delta|b|^{2}}{2}\int\frac{d^{4}k}{(2\pi)^{4}}F(k)\int d^{4}\theta\theta^{2}\bar{\theta}^{2}+h.c. \nonumber\\
&=&\frac{\delta b^{2}}{\kappa}\frac{1}{\epsilon}-\frac{\delta b}{2\kappa}\left(\eta^{+}\overline{\ln}\eta^{+}-\eta^{-}\overline{\ln}\eta^{-}-2b\right) \ ; \label{G2 delta1}
\end{eqnarray}
\begin{eqnarray}
{\mathcal G}_{3}^{(1)\delta^{1}}&=&4\delta g\int\frac{d^{4}kd^{4}\theta_{12}}{(2\pi)^4}\Phi_{1}(1)\delta^{4}_{12}\left[-\frac{1}{4}\bar{D}_{1}^{2}(k)\langle\Phi_{0}\Phi_{1}\rangle\right]+h.c. \nonumber\\
&=&4\delta g\bar{a}|b|^{2}\int\frac{d^{4}k}{(2\pi)^{4}}B(k)\int d^{4}\theta\theta^{2}\bar{\theta}^{2}\Phi_{1}+h.c. \nonumber\\
&=&-\frac{4\delta ga\langle\varphi_{1}\rangle}{\kappa}\left(2\eta^{2}\overline{\ln}\eta^{2}-\eta^{+}\overline{\ln}\eta^{+}-\eta^{-}\overline{\ln}\eta^{-}\right) \ ; \label{G3 delta1}
\end{eqnarray}
\begin{eqnarray}
{\mathcal G}_{4}^{(1)\delta^{1}}&=&-\delta a\int\frac{d^{4}kd^{4}\theta_{12}}{(2\pi)^4}\delta^{4}_{12}\left[-\frac{1}{4}\bar{D}_{1}^{2}(k)\langle\Phi_{0}\Phi_{1}\rangle\right]+h.c. \nonumber\\
&=&-\delta|a|^{2}|b|^{2}\int\frac{d^{4}k}{(2\pi)^{4}}B(k)\int d^{4}\theta\theta^{2}\bar{\theta}^{2}+h.c. \nonumber\\
&=&\frac{\delta a^{2}}{\kappa}\left(2\eta^{2}\overline{\ln}\eta^{2}-\eta^{+}\overline{\ln}\eta^{+}-\eta^{-}\overline{\ln}\eta^{-}\right) \ ; \label{G4 delta1}
\end{eqnarray}
\begin{eqnarray}
{\mathcal G}_{5}^{(1)\delta^{1}}&=&-\delta\rho\int\frac{d^{4}kd^{4}\theta_{12}}{(2\pi)^4}\delta^{4}_{12}\left[-\frac{1}{4}\bar{D}_{1}^{2}(k)\langle\Phi_{1}\Phi_{2}\rangle\right]+h.c. \nonumber\\
&=&-\delta\rho\bar{M}|b|^{2}\int\frac{d^{4}k}{(2\pi)^{4}}B(k)\int d^{4}\theta\theta^{2}\bar{\theta}^{2}+h.c. \nonumber\\
&=&\frac{\delta\rho M}{\kappa}\left(2\eta^{2}\overline{\ln}\eta^{2}-\eta^{+}\overline{\ln}\eta^{+}-\eta^{-}\overline{\ln}\eta^{-}\right) \ , \label{G5 delta1}
\end{eqnarray}
with  $\Phi_{0}(1)=\Phi_{0}(p=0, \theta_{1},\bar{\theta}_{1})$.
 
Let us deal with the divergent ones, ${\mathcal G}_{1}^{(1)\delta^{1}}$ and ${\mathcal G}_{2}^{(1)\delta^{1}}$, and apply the renormalization prodecure. We are going to use the $\overline{MS}$ scheme. The diagram  ${\mathcal G}_{2}^{(1)\delta^{1}}$ is a vacuum diagram, so its renormalization is trivial. We just need to cancel the divergence with a constant counterterm, which implies a redefinition of the vacuum energy. Since  the tadpole diagrams are not identically zero, the interpolated theory has a new divergence, which apparently is not canceled by a counterterm of the K\"{ahlerian} type potential. To renormalize the divergence in ${\mathcal G}_{1}^{(1)\delta^{1}}$, we need the counterterm
\begin{equation}
\frac{2\delta gb}{\kappa\epsilon}\int d^{4}\theta\bar{\theta}^{2}\Phi_{0R}+h.c.=\frac{2\delta gb}{\kappa\epsilon}\int d^{2}\theta\Phi_{0R}+h.c. \ , \label{CT1}
\end{equation}
where $\Phi_{0R}$ denotes the renormalized superfield.  It seems that the quiral potential is renormalized, which is strange because the renormalization structure of the theory should not be modified by the soft terms introduced by the method. However, this counterterm depends on $b$, wich must be a solution of the optimization procedure. 

The renormalized effective potential up to the order $\delta^1$ is  
\begin{eqnarray}
{\mathcal V}_{eff}^{(1)}&=&\frac{1}{(4\pi)^{2}}\left\{\frac{1}{4}\!\left(M^{2}\!+\!a^{2}\right)^{2}\ln\!\left[1-\frac{b^{2}}{\left(M^{2}+a^{2}\right)^{2}}\right]\right. \nonumber\\
&&\left.\hspace{1,5cm}+\frac{b}{2}\!\left(M^{2}\!+\!a^{2}\right)\!\ln\!\left[\frac{M^{2}+a^{2}+b}{M^{2}+a^{2}-b}\right]\!+\!\frac{b^{2}}{4}\ln\!\left[\frac{\left(M^{2}\!+\!a^{2}\right)^{2}-b^{2}}{\mu^4}\right]\!-\!\frac{3b^2}{4}\right\} \nonumber\\
&&+\frac{\delta}{(4\pi)^{2}}\!\left\{b(b-4g\langle F_{0}\rangle)+2\left[a(a-4g\langle\varphi_{1}\rangle)+\rho M\right]\!\left(M^{2}+a^{2}\right)\overline{\ln}\left[M^{2}+a^{2}\right]\right. \nonumber\\
&&\!\left.+\left[a(4g\langle\varphi_{1}\rangle\!-\!a)\!+\!\frac{1}{2}(4g\langle F_{0}\rangle\!-\!b)\!-\!\rho M\right]\!\!\left(M^{2}\!+\!a^{2}\!+\!b\right)\overline{\ln}\left[M^{2}\!+\!a^{2}\!+\!b\right]\right. \nonumber\\
&&\!\left.+\left[a(4g\langle\varphi_{1}\rangle\!-\!a)\!-\!\frac{1}{2}(4g\langle F_{0}\rangle\!-\!b)\!-\!\rho M\right]\!\!\left(M^{2}\!+\!a^{2}\!-\!b\right)\overline{\ln}\left[M^{2}\!+\!a^{2}\!-\!b\right]\!\right\}. \label{1loopdelta1}
\end{eqnarray}

At this stage, we have a perturbative result for the effective potential. In order to get a nonperturbative result, we apply the optimization procedure. Since we split the parameters $M_{ij}$ into a $\theta$-independent ($a_{ij}$) and a $\theta$-dependent ($b_{ij}$) part, and recalling (\ref{choice}), the optimized parameters will be $a_{01}=a$, $b_{11}=b$, and $\rho_{12}=\rho$. Using the \textit{PMS} criterion to find the optimized parameters $a$, $b$ and $\rho$, we have to solve the three coupled equations
\begin{equation}
\left.\frac{\partial{\mathcal V}_{eff}^{\delta^2}}{\partial a}\right|_{a=a_0}=\left.\frac{\partial{\mathcal V}_{eff}^{\delta^2}}{\partial b}\right|_{b=b_0}=\left.\frac{\partial{\mathcal V}_{eff}^{\delta^2}}{\partial\rho}\right|_{\rho=\rho_0}=0 \ , \label{PMS OR 2 Veff}
\end{equation}
at $\delta=1$, and plug the optimized values $a_0$, $b_0$ and $\rho_0$ into (\ref{1loopdelta1}). We find the following analytical solutions:
\begin{eqnarray}
a_{0}&=&4g\langle\varphi_{1}\rangle=\bar{a_{0}} \nonumber\\
b_{0}&=&4g\langle F_{0}\rangle=\bar{b_{0}} \ , \nonumber \\
\rho_{0}&=&0=\bar{\rho_{0}} \ . \label{analiticalsol}
\end{eqnarray}

This result shows that the optimized parameters  are functions of the original coupling and fields, as we expected. It is easy to see that this is also a solution of the {\it FAC} criterion. However, as we have three optimized parameters, it is not clear how we may write three optimization equations using this criterion. Replacing these values in (\ref{pot efet OR ordem 1}), all the $\delta^1$ terms vanish and the optimized potential is written as:
\begin{eqnarray}
{\mathcal V}_{eff}^{(1)}={\mathcal G}^{(1)\delta^0}&=&\frac{1}{(8\pi)^{2}}\left\{(m^{2}+16g^{2}\langle\varphi_{1}\rangle^{2})^{2}\ln\left[1-\frac{16g^{2}\langle F_{0}\rangle^{2}}{(m^{2}+16g^{2}\langle\varphi_{1}\rangle^{2})^{2}}\right]\right. \nonumber\\ 
&&\left.+8g\langle F_{0}\rangle(m^{2}+16g^{2}\langle\varphi_{1}\rangle^{2})\ln\left[\frac{m^{2}+16g^{2}\langle\varphi_{1}\rangle^{2}+4g\langle F_{0}\rangle}{m^{2}+16g^{2}\langle\varphi_{1}\rangle^{2}-4g\langle F_{0}\rangle}\right]\right. \nonumber\\
&&\left.+16g^{2}\langle F_{0}\rangle^{2}\ln\left[\frac{(m^{2}+16g^{2}\langle\varphi_{1}\rangle^{2})^{2}-16g^{2}\langle F_{0}\rangle^{2}}{\mu^4}\right]- 48g^2\langle F_0\rangle ^2\right\} \ . \label{pot Coleman Weinberg OR final}  
\end{eqnarray}

This is the Coleman-Weinberg potential for the O'Raifeartaigh model \cite{Helayel} and represents the sum of all one-loop diagrams, i.e., a nonperturbative result, because it takes into account infinite orders  of the orginal coupling constant.  
 
Let us come back to the counterterm defined in (\ref{CT1}). Since $ \Phi_0 $ and $\Phi_1$  are classical superfields, the optimized parameters can be written as:
\begin{eqnarray}
{a}_{0}&=&4g\int d^{2}\bar{\theta}\bar{\Phi}_{1}=4g\int d^{2}\theta\Phi_{1}=4g\langle\varphi_{1}\rangle , \nonumber\\
{b}_{0}&=&4g\int d^{2}\bar{\theta}\bar{\Phi}_{0}=4g\int d^{2}\theta\Phi_{0}=4g\langle F_{0}\rangle. \label{ab1loop}
\end{eqnarray}
This is in agreement with the {\it R} charges of the parameters $a$ and $b$. Thus, putting the expression above for $b_0$ into (\ref{CT1}) we see that the counterterm needed to renormalize the divergence in ${\mathcal G}_{1}^{(1)\delta^{1}}$ is of the form
\begin{equation}
\frac{8\delta g^{2}}{\kappa\epsilon}\int d^{4}\theta\bar{\Phi}_{0R}\Phi_{0R}+h.c.=\frac{16\delta g^{2}}{\kappa\epsilon}\int d^{4}\theta\bar{\Phi}_{0R}\Phi_{0R} \ , \label{CT1superspace}
\end{equation}
and we note that in fact, after the optimization procudure, only the K\"{a}hler potential is renormalized, in agreement with the nonrenormalization theorem. In the next section we discuss  the  divergences at order $\delta^2$.

\section{Order two results}

At order two we have one- and two-loop diagrams.  Let us start with the one-loop diagrams. They  have three topologies: diagrams with two external legs and no insertions, diagrams with one external leg and one insertion and vacuum diagrams with no external legs and two insertions. There are 42 such diagrams (plus the hermitian conjugates). However, since our main concern in the present work is to analyse the divergent structure of the theory, we write below only the three divergent diagrams, shown  in Fig. 3.
\newpage
\begin{eqnarray*}
\begin{picture}(330,5) \thicklines 
\put(15,0){\line(50,0){30}}\put(15,4){$\Phi_0$}\put(35,12){$\Phi_1$}\put(35,-17){$\Phi_1$}\put(61,0){\circle{30}}\put(75,12){$\bar{\Phi}_1$}\put(75,-17){$\bar{\Phi}_1$}\put(77,0){\line(50,0){30}}\put(94,4){$\bar{\Phi}_0$}\put(114,-3){;}
\put(125,0){\line(50,0){30}}\put(125,4){$\Phi_0$}\put(145,12){$\Phi_1$}\put(145,-17){$\Phi_1$}\put(171,0){\circle{30}}\put(185,12){$\bar{\Phi}_1$}\put(185,-17){$\bar{\Phi}_1$}\put(182,-3){$\otimes$}\put(192,-3){$\bar{\theta}^2$}\put(208,-3){+ $h.c.$}\put(245,-3){;}
\put(255,-3){$\theta^2$}\put(265,-3){$\times$}\put(260,12){$\Phi_1$}\put(260,-17){$\Phi_1$}\put(286,0){\circle{30}}\put(300,12){$\bar{\Phi}_1$}\put(300,-17){$\bar{\Phi}_1$}\put(297,-3){$\otimes$}\put(307,-3){$\bar{\theta}^2$}
\end{picture}
\end{eqnarray*}
\vspace{-0,3cm}
\begin{center}
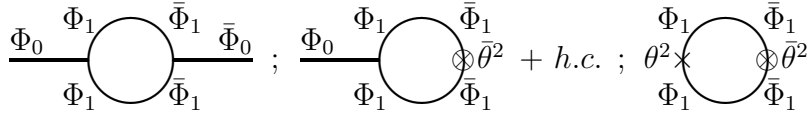
Fig. 3: Divergent diagrams ${\mathcal G}_{1}^{(1)\delta^2}$, ${\mathcal G}_{2}^{(1)\delta^2}$ and ${\mathcal G}_{3}^{(1)\delta^2}$. 
\end{center}
\begin{eqnarray}
{\mathcal G}_{1}^{(1)\delta^2}&=&2(2\delta g)(2\delta g)\int\frac{d^{4}kd^{4}\theta_{12}}{(2\pi)^{4}}\Phi_{0}(1)\bar{\Phi}_{0}(2)\left[-\frac{1}{4}\bar{D}_{1}^{2}(k)\langle\Phi_{1}\bar{\Phi}_{1}\rangle\right]\left[-\frac{1}{4}D_{2}^{2}(-k)\langle\bar{\Phi}_{1}\Phi_{1}\rangle\right] \nonumber\\
&=&\frac{8\delta^{2}g^{2}}{16}\int\frac{d^{4}k}{(2\pi)^{4}}\left\{E(k)E(k){\mathcal J}_{1}(\theta,\bar\theta)+2\frac{|b|^{2}}{16}E(k)B(k){\mathcal J}_{2}(\theta,\bar\theta)\right. \nonumber\\
&&\left.\hspace{3,5cm}+\frac{|b|^{4}}{(16)^{2}}B(k)B(k){\mathcal J}_{3}(\theta,\bar\theta)\right\} \nonumber\\
&=&\frac{8\delta^{2}g^{2}\langle F_{0}\rangle^{2}}{\kappa}\left(\frac{1}{\epsilon}-\overline{\ln}\eta^{2}\right) \nonumber\\
&&+\frac{2\delta^{2}g^{2}\langle F_{0}\rangle^{2}}{\kappa b}\left[4b\overline{\ln}\eta^{2}-(\eta^{2}+2b)\overline{\ln}\eta^{+}+(\eta^{2}-2b)\overline{\ln}\eta^{-}+2b\right] \ . \label{G1div}
\end{eqnarray}
\begin{eqnarray}
{\mathcal G}_{2}^{(1)\delta^2}&=&2(2\delta g)\left(-\frac{\delta\bar{b}}{2}\right)\int\frac{d^{4}kd^{4}\theta_{12}}{(2\pi)^{4}}\Phi_{0}(1)\bar{\theta}_{2}^{2}\left[-\frac{1}{4}\bar{D}_{1}^{2}(k)\langle\Phi_{1}\bar{\Phi}_{1}\rangle\right]\left[-\frac{1}{4}D_{2}^{2}(-k)\langle\bar{\Phi}_{1}\Phi_{1}\rangle\right]+h.c. \nonumber\\
&=&-\frac{2\delta^{2}g}{16}\bar{b}\int\frac{d^{4}k}{(2\pi)^{4}}\left\{E(k)E(k){\mathcal J}_{4}(\theta,\bar\theta)+2\frac{|b|^{2}}{16}E(k)B(k){\mathcal J}_{5}(\theta,\bar\theta)\right. \nonumber\\
&&\left.\hspace{3,5cm}+\frac{|b|^{4}}{(16)^{2}}B(k)B(k){\mathcal J}_{6}(\theta,\bar\theta)\right\}+h.c. \nonumber\\
&=&-\frac{4\delta^{2}gb\langle F_{0}\rangle}{\kappa}\left(\frac{1}{\epsilon}-\overline{\ln}\eta^{2}\right) \nonumber\\
&&-\frac{\delta^{2}gb\langle F_{0}\rangle}{\kappa b}\left[4b\overline{\ln}\eta^{2}-(\eta^{2}+2b)\overline{\ln}\eta^{+}+(\eta^{2}-2b)\overline{\ln}\eta^{-}+2b\right] \ . \label{G2div}
\end{eqnarray}
\begin{eqnarray}
{\mathcal G}_{3}^{(1)\delta^2}&=&2\left(\frac{\delta b}{2}\right)\left(\frac{\delta\bar{b}}{2}\right)\int\frac{d^{4}kd^{4}\theta_{12}}{(2\pi)^{4}}\theta_{1}^{2}\bar{\theta}_{2}^{2}\left[-\frac{1}{4}\bar{D}_{1}^{2}(k)\langle\Phi_{1}\bar{\Phi}_{1}\rangle\right]\left[-\frac{1}{4}D_{2}^{2}(-k)\langle\bar{\Phi}_{1}\Phi_{1}\rangle\right] \nonumber\\
&=&\frac{\delta^{2}}{2(16)}|b|^{2}\int\frac{d^{4}k}{(2\pi)^{4}}\left\{E(k)E(k){\mathcal J}_{7}(\theta,\bar\theta)+2\frac{|b|^{2}}{16}E(k)B(k){\mathcal J}_{8}(\theta,\bar\theta)\right. \nonumber\\
&&\left.\hspace{3,5cm}+\frac{|b|^{4}}{(16)^{2}}B(k)B(k){\mathcal J}_{9}(\theta,\bar\theta)\right\} \nonumber\\
&=&\frac{\delta^{2}b^{2}}{2\kappa}\left(\frac{1}{\epsilon}-\overline{\ln}\eta^{2}\right) \nonumber\\
&&+\frac{\delta^{2}b^{2}}{8\kappa b}\left[4b\overline{\ln}\eta^{2}-(\eta^{2}+2b)\overline{\ln}\eta^{+}+(\eta^{2}-2b)\overline{\ln}\eta^{-}+2b\right] \ . \label{G3div}
\end{eqnarray} 

There are four two-loop diagrams (plus the hermitian conjugates). They are all vacuum diagrams and differ from each other by the propagators appearing in the loops. Out of these four diagrams, only one gives a divergent contribution, and we show it in Fig. 4.
\newpage
\begin{eqnarray*}
\begin{picture}(375,5) \thicklines 
\put(188,0){\circle{40}}\put(188,-20){\line(0,50){40}}\put(173,24){$\bar{\Phi}_1$}\put(173,-29){$\Phi_1$}\put(190,24){$\bar{\Phi}_1$}\put(192,-29){$\Phi_1$}\put(174,-12){$\Phi_0$}\put(189,7){$\bar{\Phi}_0$}
\end{picture}
\end{eqnarray*}
\vspace{-0,3cm}
\begin{center}
Fig. 4: Divergent diagram ${\mathcal G}_{1}^{(2)\delta^2}$.
\end{center} 

As the one-loop diagram ${\mathcal G}_{3}^{(1)\delta^2}$, this is a vacuum diagram, and its renormalization is trivial. Again, we just need to cancel the divergence with a constant counterterm, which implies a redefinition of the vacuum energy. 

To renormalize the divergent term in ${\mathcal G}_{1}^{(1)\delta^2}$ we introduce the counterterm
\begin{equation}
-\frac{8\delta^{2}g^{2}}{\kappa\epsilon}\int d^{4}\theta\Phi_{0R}\bar{\Phi}_{0R} \ , \label{CT2}
\end{equation}
and, for the divergent term in ${\mathcal G}_{2}^{(1)\delta^2}$, we introduce the counterterm
\begin{equation}
\frac{2\delta^{2}g}{\kappa\epsilon}\int d^{4}\theta\bar{\theta}^{2}b\Phi_{0R}+h.c.
\end{equation}

As in the previous section, when we renormalized the effective potential up to the order $\delta^1$, the counterterm introduced for the divergent term in ${\mathcal G}_{2}^{(1)\delta^2}$ is proportional to $b$, and, at one-loop, the optimized parameter $b_0$ is given by Eq. (\ref{ab1loop}). 
It can be shown that this solution is valid for all orders in $\delta$ at one-loop level. Thus, the counterterm is proportional to 
\begin{equation}
\delta^{2}\int d^{4}\theta\Phi_{0R}\bar{\Phi}_{0R} \ , \label{CTdelta}
\end{equation} 
showing that in fact, at one-loop, only the wave function is renormalized. However, owing to the two-loop vacuum diagrams, for the effective potential up to the order $\delta^2$, the optimized parameters defined in (\ref{analiticalsol}) are no longer soultion to the {\it PMS} equations. In fact, when  diagrams like the one in Fig. 4 are  taken into account, we can derive two-loop nonperturbative  corrections to the Coleman-Weinberg potential. To this end it is necessary to evaluate all the order $\delta^2$  diagrams and solve numerically a complicate set of equations. This is a work in progress \cite{nos2}. Here, we are interested in the renormalization structure of the theory. Since the counterterms depend only on $b$, let us  concentrate on this parameter.
Althought it is not possible to find an analytical solution for the $b$ parameter at order $\delta^2$, we can write  the general form for it. Based on {\it R} charge considerations, on the fact that $b_0$ must be a function of the original component fields and on the Lagrangian dependence on $b$, we can argue that the optimized parameter up to the order $\delta^2$ must be 
\begin{equation}
b_0= \int d^{2}\bar{\theta}^{2} [4g\bar{\Phi}_{0}+\bar{A}_{2-loop}] \ , 
\end{equation}
where $\bar{A}_{2-loop}$ is a two-loop corretion with {\it R} charge equals to $-2$. Since all the counterterms are proportional to $\int d^{4}\theta\Phi_{0R}\bar{\Phi}_{0R}$ or $\int d^{2}\theta b\Phi_{0R}$, it can be seen that only the K\"{a}hler potential is renormalized.  


\section{Concluding Remarks}

Our efforts in the present work have been focused on the application of superfield techniques and supergraph calculations to study the renormalization of the minimal O'Raifeartaigh model in the LDE scheme to the second order in the expansion parameter. Our calculations show that only the K\"{a}hler potential gets renormalized, according to what should be expected from the ${\mathcal N}=1, \ D=4$ SUSY nonrenormalization theorem for the chiral potential. We point out that, in this paper, we are actually interested in understanding and mastering the superfield approach if we adopt the LDE procedure to compute higher order corrections to the effective potential in the case of spontaneously broken SUSY.  

Here, we have not yet concentrated on the task of effective potential calculation. As already stated in the Introduction, this is the matter of a forthcoming work. Our main purpose in the present paper was to check the consistency and the efficacy of superfield and supergraph methods to deal with SUSY explicitly breaking terms in a higher order loop computation in superspace. We have checked, with our explicit supergraph computations, that the extended super-Feynman rules are perfectly consistent even though these explicitly breaking terms show up. The structure of divergences has suitably been treated and the final result of the K\"{a}hler potential renormalization is a good check of our manipulations. Once the renormalization task is accomplished, it remains to be done - and we believe this requires a forthcoming work  - the complete two-loop calculation to allow us to proceed to the next step, namely the optimization in the LDE parameter, to finally write down the full two-loop corrected effective potential. This demands a nontrivial work in terms of Feynman supergraph computation and numerial computation and we shall be soon reporting on our results \cite{nos2}.

Once this whole program of higher order corrected effective potentials has been accomplished for the F-term SUSY breaking, we believe it would be worthwhile to concentrate efforts on the LDE effective potential calculation in the case of D-term SUSY breakings in the gauge sector and to pay attention to the problem of metastable SUSY breaking vacua and its connection with {\it R} symmetry spontaneous breaking, which has direct consequences to the physics of the so-called lightest supersymmetric particle.


\section{Acknowledgements}

J. A. Helay\"{e}l-Neto expresses his gratitude to FAPERJ-RJ for the financial support. Daniel L. Nedel would like to thank CNPq, grant 501317/2009-0, for financial support. Carlos R. Senise Jr. thanks CAPES-Brazil and Programa Rec\'em-Doutor-UNESP for financial support.  


\section*{Appendix A: Superspace integrals}
This Appendix is devoted to the evaluation of some of the superspace one-loop integrals which arise in order $\delta^2$. The integrals that appear in the expressions for the divergent one-loop diagrams of order $\delta^2$ are:
\begin{eqnarray}
{\mathcal J}_{1}(\theta,\bar\theta)&=&\int d^{4}\theta_{12}\Phi_{0}(1)\bar{\Phi}_{0}(2)\left[\bar{D}_{1}^{2}(k)\delta^{4}_{12}\right]\left[D_{2}^{2}(-k)\delta^{4}_{12}\right] \nonumber\\
&=&16\langle F_{0}\rangle^{2} \ ; \\
&& \nonumber\\
{\mathcal J}_{2}(\theta,\bar\theta)&=&\int d^{4}\theta_{12}\Phi_{0}(1)\bar{\Phi}_{0}(2)\left[\bar{D}_{1}^{2}(k)\delta^{4}_{12}\right]\left[D_{2}^{2}(-k)\bar{D}_{2}^{2}(-k)\bar{\theta}_{2}^{2}\theta_{2}^{2}D_{2}^{2}(-k)\delta^{4}_{12}\right] \nonumber\\
&=&(16)^{2}\langle F_{0}\rangle^{2} \ ; \\
&& \nonumber\\
{\mathcal J}_{3}(\theta,\bar\theta)&=&\int d^{4}\theta_{12}\Phi_{0}(1)\bar{\Phi}_{0}(2)\left[\bar{D}_{1}^{2}(k)D_{1}^{2}(k)\theta_{1}^{2}\bar{\theta}_{1}^{2}\bar{D}_{1}^{2}(k)\delta^{4}_{12}\right]\left[D_{2}^{2}(-k)\bar{D}_{2}^{2}(-k)\bar{\theta}_{2}^{2}\theta_{2}^{2}D_{2}^{2}(-k)\delta^{4}_{12}\right] \nonumber\\
&=&(16)^{3}\langle F_{0}\rangle^{2} \ ; \\
&& \nonumber\\
{\mathcal J}_{4}(\theta,\bar\theta)&=&\int d^{4}\theta_{12}\Phi_{0}(1)\left[\bar{D}_{1}^{2}(k)\delta^{4}_{12}\right]\left[\bar{\theta}_{2}^{2}D_{2}^{2}(-k)\delta^{4}_{12}\right] \nonumber\\
&=&16\langle F_{0}\rangle \ ; \\
&& \nonumber\\
{\mathcal J}_{5}(\theta,\bar\theta)&=&\int d^{4}\theta_{12}\Phi_{0}(1)\left[\bar{D}_{1}^{2}(k)\bar{\theta}_{1}^{2}D_{1}^{2}(k)\delta^{4}_{12}\right]\left[\theta_{2}^{2}\bar{D}_{2}^{2}(-k)\bar{\theta}_{2}^{2}D_{2}^{2}(-k)\delta^{4}_{12}\right] \nonumber\\
&=&(16)^{2}\langle F_{0}\rangle \ ; \\
&& \nonumber\\
{\mathcal J}_{6}(\theta,\bar\theta)&=&\int d^{4}\theta_{12}\Phi_{0}(1)\left[\bar{D}_{1}^{2}(k)D_{1}^{2}(k)\theta_{1}^{2}\bar{\theta}_{1}^{2}\bar{D}_{1}^{2}(k)\delta^{4}_{12}\right]\left[\bar{\theta}_{2}^{2}D_{2}^{2}(-k)\bar{D}_{2}^{2}(-k)\bar{\theta}_{2}^{2}\theta_{2}^{2}D_{2}^{2}(-k)\delta^{4}_{12}\right] \nonumber\\
&=&(16)^{3}\langle F_{0}\rangle \ ; \\
&& \nonumber\\
{\mathcal J}_{7}(\theta,\bar\theta)&=&\int d^{4}\theta_{12}\left[\theta_{1}^{2}\bar{D}_{1}^{2}(k)\delta^{4}_{12}\right]\left[\bar{\theta}_{2}^{2}D_{2}^{2}(-k)\delta^{4}_{12}\right] \nonumber\\
&=&16 \ ; \\
&& \nonumber\\
{\mathcal J}_{8}(\theta,\bar\theta)&=&\int d^{4}\theta_{12}\left[\theta_{1}^{2}\bar{D}_{1}^{2}(k)\bar{\theta}_{1}^{2}D_{1}^{2}(k)\delta^{4}_{12}\right]\left[\theta_{2}^{2}\bar{D}_{2}^{2}(-k)\bar{\theta}_{2}^{2}D_{2}^{2}(-k)\delta^{4}_{12}\right] \nonumber\\
&=&(16)^{2} \ ; \\
&& \nonumber\\
{\mathcal J}_{9}(\theta,\bar\theta)&=&\int d^{4}\theta_{12}\left[\theta_{1}^{2}\bar{D}_{1}^{2}(k)D_{1}^{2}(k)\theta_{1}^{2}\bar{\theta}_{1}^{2}\bar{D}_{1}^{2}(k)\delta^{4}_{12}\right]\left[\bar{\theta}_{2}^{2}D_{2}^{2}(-k)\bar{D}_{2}^{2}(-k)\bar{\theta}_{2}^{2}\theta_{2}^{2}D_{2}^{2}(-k)\delta^{4}_{12}\right] \nonumber\\
&=&(16)^{3} \ .
\end{eqnarray}  


\section*{Appendix B: Momentum space integrals}

In this Appendix we collect the expressions for the one-loop integrals, which are not so easy to obtain directly. We use dimensional regularization.

We begin with the simple integrals \cite{Nibbelink}:
\begin{eqnarray}
J(m^2)&=&\frac{(\mu^2)^{\epsilon}}{(2\pi)^D}\int\frac{d^{D}k}{k^2+m^2}=\frac{(\mu^2)^{\epsilon}}{(4\pi)^{2-\epsilon}}\Gamma\left(-1+\epsilon\right)(m^2)^{1-\epsilon} \nonumber \\ 
J_n(m^2)&=&\int\frac{d^{D}k}{(2\pi)^{D}\mu^{D-4}}\frac{1}{k^2}\frac{1}{(k^2+m^2)}
=\frac{(m^2)^{1-n}}{16\pi^2}\left(4\pi\frac{\mu^2}{m^2}\right)^{2-(D/2)}\frac{\pi}{\Gamma(D/2)\sin\pi(D/2-n)} \nonumber \\
L_n(m^{2})&=&\int\frac{d^{D}k}{(2\pi)^{D}\mu^{D-4}}\frac{1}{k^{2n}}\ln\left(1+\frac{m^2}{k^2}\right) =\frac{m^2}{D/2-n}J_n(m^{2})\
\ , \label{int JnLn}
\end{eqnarray}
where $D=4-2\epsilon$ and $\mu$ is the renormalization scale. Expanding these expressions, we obtain
the useful equalities:
\begin{equation}
J(m^2)= \frac{1}{16\pi^2}\left[-\frac{m^2}{\epsilon}+m^2\left(\ln \frac{m^2}{\mu^2}-1\right)\right] \ , \label{J(x) 3}
\end{equation}
\begin{equation}
J_0(m^2)=-\frac{m^2}{16\pi^2}\left[\frac{1}{\epsilon}+1-\ln\frac{m^2}{\mu^2}\right] \ , \label{int J0}
\end{equation}
\begin{equation}
J_1(m^2)=\frac{1}{16\pi^2}\left[\frac{1}{\epsilon}+1-\ln\frac{m^2}{\mu^2}\right] \ , \label{int J1}
\end{equation}
\begin{equation}
L_0(m^2)=-\frac{1}{2}\frac{m^4}{16\pi^2}\left[\frac{1}{\epsilon}+\frac{3}{2}-\ln\frac{m^2}{\mu^2}\right] \ , \label{int L0}
\end{equation}
\begin{equation}
L_1(m^2)=\frac{m^2}{16\pi^2}\left[\frac{1}{\epsilon}+2-\ln\frac{m^2}{\mu^2}\right] \ . \label{int L1}
\end{equation}

In the calculation of the order $\delta^0$ diagram, we have to solve the integral
\begin{equation}
K(m^2,M^2)=\int\frac{d^{D}k}{(2\pi)^{D}\mu^{D-4}}tr\ln\left(1-\frac{M^2\bar{M}^2}{(k^2+m^2\bar{m}^2)^2}\right) \label{K mM} \ .
\end{equation}
which can be written as a sum of integrals of the type (\ref{int L0}): 
\begin{equation}
K(m^2,M^2)=tr\left[L_0(m^2+\tilde{m}^2)+L_0(m^2-\tilde{m}^2)-2L_0(m^2)\right] \ , \label{K mM soma}
\end{equation}
with $\tilde{m}^{2}=\left(M^2\bar{M}^2\right)^{1/2}$.

In the following, we define 
\begin{equation}
\eta^{2}=M^{2}+a^{2} \ \ \ , \ \ \ \eta^{\pm}=M^{2}+a^{2}\pm b \ , \label{etas} 
\end{equation}
and we adopt the same notation of \cite{Jones,Espinosa,Martin}.

Using the definition of the propagators we now show the results for the integrals appearing in the one-loop diagrams of orders $\delta^1$ and $\delta^2$. To solve them we use the method of partial fraction, splitting each integral as a sum of other integrals with just one propagator in the integrand. Doing this, we obtain:
\begin{eqnarray}
I_{1}(k)&=&\int\frac{d^{4}k}{(2\pi)^4}F(k) \nonumber\\
&=&\int\frac{d^{4}k}{(2\pi)^4}\frac{1}{\left(k^{2}+M^{2}+a^{2}\right)^{2}-b^{2}} \nonumber\\
&=&-\frac{1}{2b}J(\eta^+)+\frac{1}{2b}J(\eta^-) \nonumber\\
&=&\frac{1}{\kappa\epsilon}-\frac{1}{2\kappa b}\left(\eta^{+}\overline{\ln}\eta^{+}-\eta^{-}\overline{\ln}\eta^{-}-2b\right) \ , \\
&& \nonumber\\
I_{2}(k)&=&\int\frac{d^{4}k}{(2\pi)^4}B(k) \nonumber\\
&=&\int\frac{d^{4}k}{(2\pi)^4}\frac{1}{\left(k^{2}+M^{2}+a^{2}\right)\left[\left(k^{2}+M^{2}+a^{2}\right)^{2}-b^{2}\right]} \nonumber\\
&=&-\frac{1}{b^2}J(\eta^2)+\frac{1}{2b^2}J(\eta^+)+\frac{1}{2b^2}J(\eta^-) \nonumber\\
&=&-\frac{1}{2\kappa b^2}\left(2\eta^{2}\overline{\ln}\eta^{2}-\eta^{+}\overline{\ln}\eta^{+}-\eta^{-}\overline{\ln}\eta^{-}\right) \ , \\
&& \nonumber\\
I_{3}(k)&=&\int\frac{d^{4}k}{(2\pi)^4}E(k)E(k) \nonumber\\
&=&\int\frac{d^{4}k}{(2\pi)^4}\frac{1}{\left(k^{2}+M^{2}+a^{2}\right)\left(k^{2}+M^{2}+a^{2}\right)} \nonumber\\
&=&\tilde{J}(\eta^2) \nonumber\\
&=&\frac{1}{\kappa\epsilon}-\frac{1}{\kappa}\overline{\ln}\eta^2 \ , \\
&& \nonumber\\
I_{4}(k)&=&\int\frac{d^{4}k}{(2\pi)^4}B(k)E(k) \nonumber\\
&=&\int\frac{d^{4}k}{(2\pi)^4}\frac{1}{\left(k^{2}+M^{2}+a^{2}\right)\left[\left(k^{2}+M^{2}+a^{2}\right)^{2}-b^{2}\right]\left(k^{2}+M^{2}+a^{2}\right)} \nonumber\\
&=&\frac{1}{b^4}\tilde{J}(\eta^2)+\frac{1}{4b^4}\tilde{J}(\eta^+)+\frac{1}{4b^4}\tilde{J}(\eta^-)+\frac{3}{4b^5}J(\eta^+)-\frac{3}{4b^5}J(\eta^-) \nonumber\\
&=&\frac{1}{2\kappa b^3}\left(2b\overline{\ln}\eta^{2}-\eta^{+}\overline{\ln}\eta^{+}+\eta^{-}\overline{\ln}\eta^{-}+2b\right) \ , \\ 
&& \nonumber\\
I_{5}(k)&=&\int\frac{d^{4}k}{(2\pi)^4}B(k)B(k) \nonumber\\
&=&\int\frac{d^{4}k}{(2\pi)^4}\frac{1}{\left(k^{2}+M^{2}+a^{2}\right)\left[\left(k^{2}+M^{2}+a^{2}\right)^{2}-b^{2}\right]\left(k^{2}+M^{2}+a^{2}\right)\left[\left(k^{2}+M^{2}+a^{2}\right)^{2}-b^{2}\right]} \nonumber\\
&=&\frac{1}{b^4}\tilde{J}(\eta^2)+\frac{1}{4b^4}\tilde{J}(\eta^+)+\frac{1}{4b^4}\tilde{J}(\eta^-)+\frac{3}{4b^5}J(\eta^+)-\frac{3}{4b^5}J(\eta^-) \nonumber\\
&=&-\frac{1}{4\kappa b^5}\left[4b\overline{\ln}\eta^{2}-(3\eta^{+}-b)\overline{\ln}\eta^{+}+(3\eta^{-}+b)\overline{\ln}\eta^{-}+6b\right] \ .
\end{eqnarray}


\end{document}